\def\ANON{0}

\documentclass{llncs}
\usepackage{times}
\usepackage{epsfig}
\usepackage{graphicx}
\usepackage{amsmath}
\usepackage{amssymb}

\usepackage{array}
\usepackage[hyphens]{url}
\usepackage{caption}
\usepackage{graphicx}
\usepackage{multirow}
\usepackage{amsmath}
\usepackage{multirow}
\usepackage{color}
\usepackage{xcolor}
\usepackage{algorithm}
\usepackage{algpseudocode}
\usepackage{amsfonts}
\usepackage{float}
\usepackage{subcaption}
\usepackage[pagebackref]{hyperref}
\usepackage[noabbrev]{cleveref}

\newcommand\Algphase[1]{%
\vspace*{-.1\baselineskip}\Statex\hspace*{\dimexpr-\algorithmicindent-2pt\relax}\
\Statex\hspace*{-\algorithmicindent}\textbf{#1}%
\vspace*{-.7\baselineskip}\Statex\hspace*{\dimexpr-\algorithmicindent-2pt\relax}\rule{\textwidth}{0.4pt}%
}

\newcommand{\ignore}[1]{}

\begin{document}

\title{Practical Privacy-Preserving Identity Verification\\ using Third-Party Cloud Services and FHE}
\subtitle{Role of Data Encoding in Circuit Depth Management}

\ifnum\ANON=0

\author{Deep Inder Mohan \and Srinivas Vivek\\
{\tt\small \{deepinder.mohan, srinivas.vivek\}@iiitb.ac.in}}
\institute{IIIT Bangalore, India}


\else
\author{ }
\authorrunning{ }
\institute{ }
\email{ }
\fi

\maketitle
\pagestyle{plain}

\begin{abstract}
National digital identity verification systems have played a critical role in the effective distribution of goods and services, particularly, in developing countries. Due to the cost involved in deploying and maintaining such systems, combined with a lack of in-house technical expertise, governments seek to outsource this service to third-party cloud service providers to the extent possible. This leads to increased concerns regarding the privacy of users' personal data. In this work, we propose a practical privacy-preserving digital identity (ID) verification protocol where the third-party cloud services process the identity data encrypted using a (single-key) Fully Homomorphic Encryption (FHE) scheme such as BFV. Though the role of a trusted entity such as government is not completely eliminated, our protocol does significantly reduces the computation load on such parties. 
 
A challenge in implementing a privacy-preserving ID verification protocol using FHE is to support various types of queries such as exact and/or fuzzy demographic and biometric matches including secure age comparisons. From a cryptographic engineering perspective, our main technical contribution is a user data encoding scheme that encodes demographic and biometric user data in only two BFV ciphertexts and yet facilitates us to outsource various types of ID verification queries to a third-party cloud. Our encoding scheme also ensures that the only computation done by the trusted entity is a query-agnostic ``extended'' decryption. This is in stark contrast with recent works that outsource all the non-arithmetic operations to a trusted server. We implement our protocol using the Microsoft SEAL FHE library and demonstrate its practicality.   
               
\end{abstract}



\section{Introduction}

A country's identification (ID) system plays a critical role in effective delivery of public and private services. Governments are exploring the development of multipurpose foundational digital ID systems where its individuals receive a unique identifier that they can use for identity verification. Over the last decade, some developing nations have pioneered their own digital identity verification systems, such as India with AADHAR \cite{aadhar}. 

The success of these systems have brought to  forefront the impact that national digital identity verification systems that use biometric verification can have on empowering the poor and most vulnerable sections of society; from the distribution of rations at subsidised rates, to enabling adults to find employment through employment guarantee schemes, etc. Digital identity verification systems can ensure transparent and equitable distribution of the funds and benefits from government social welfare programmes and thus help the governments of developing economies utilise their scarce development funds more effectively and efficiently. Systems such as AADHAR have become strategic policy tools for social and financial inclusion, public sector delivery reforms, managing fiscal budgets, increasing convenience and promoting hassle-free people-centric governance.

A nation deploying a digital ID system must spend considerable resources for its design, deployment and maintenance. Many developing countries may not be able to afford the design of such systems from scratch. Motivated by these needs, the Modular Open Source Identity Platform (MOSIP) project \cite{mosipwebsite} 
assists governments and other user organisations in implementing a digital, foundational identity system in a cost effective way. The Governments of Morocco, Philippines, and Ethiopia have adopted this platform, and many countries across Asia, Africa, and Latin America have expressed interest \cite{mosipnews}. 
Currently, about 84 million users have been enrolled in MOSIP-based systems \cite{mosipwebsite}.     

To ensure the availability of ID services and to reduce the burden of maintenance on central authorities, users' demographic and biometric data are often outsourced to third parties. The number of third-party cloud services involved depends on the usage of the ID service by service providers and the latter could bear the cost of these cloud services, thereby, making it economically viable for central authorities. However, this leads to serious privacy concerns and potential misuse of personal data 
\cite{aadharleak}. 
In this work, we aim to provide a practical solution to the above privacy concern by making use of fully homomorphic encryption to provide confidentiality for the demographic and biometric data outsourced to a Third-Party Server(s) (TPS). 

\noindent\textbf{Related Works}.
There are several recent works that focus on privacy-preserving biometric matching including fingerprint, facial, and iris matching. 
These works employ both cryptographic and non-cryptographic primitives to achieve privacy. See \cite[Section 2]{EngelsmaJB22} for a recent
overview of literature on privacy-preserving biometric matching. In particular, the recent works \cite{1,6,GOMEZBARRERO2017149,ppbm,BassitHVP22,Boddeti18,SperlingRRB22,EngelsmaJB22} specially make use of homomorphic encryption to achieve privacy. 

Needless to say, an ID system must support other types of queries on demographic 
information such as exact name match, date of birth, address, gender match, and also 
comparison queries such as \textit{age greater than 18?}, etc. When performing these queries on 
demographic data, we often need the result to be accurate and, hence, fuzzy matching techniques 
employed for biometric match may turn out be inadequate for demographic ID verification. While a 
trivial solution is to simply integrate different privacy-preserving protocols for demographic and biometric 
matching, such solutions do not facilitate a compact representation of user data, hence, leading to a 
relatively large usage of memory and run time. Hence, there is a need to specially design privacy-preserving ID verification protocols 
that support various types of ID verification queries. Further, 
our goal is to outsource as much computation as possible to TPS during ID verification. However, due to high 
computational complexity of FHE schemes, it is currently infeasible to outsource the entire computation to TPS. In 
all the works cited above, only the arithmetic operations (such as squared Euclidean distance computations) are performed at TPS using FHE schemes while the (high 
multiplicative depth) logical operations (such as comparisons) are performed on plaintext after decryption. It currently remains an open 
problem as to how to efficiently outsource all computations to TPS except, of course, the decryption operation while 
using FHE schemes for protecting the privacy of users' personal data.    

There are several existing identity verification systems that support various types of ID verification queries. See \cite{GunasingheKBKCS19,olympus,LeeCOK21} and references therein. But none of them facilitate the use of third-party cloud services to the best of our knowledge. 

Our choice to go with FHE schemes is motivated by the need to design a scalable system that can handle a large variety of queries in the future.
While it is possible to perform exact match, 
for example, exact name match, using additive Homomorphic Encryption (HE) schemes, the main drawback is that additive HE schemes can only perform linear operations. Hence, the input ID data to be verified must be in plaintext format for non-exact matching queries such as those involving the comparison operation. Hence, such protocols do not facilitate the use of TPS to full extent.

Also, in our setting the number of Service Providers (SP) who avail the digital ID verification service can be very large, so it is currently not realistic to make use of Multi-Key FHE (MKFHE) primitive \cite{Lopez-AltTV12} that allow an SP itself to decrypt the encrypted result of identity verification. Apart from the efficiency issues of current MKFHE schemes, there are also concerns regarding enabling an SP to decrypt ciphertexts as it could potentially encourage nexus with TPS to leak users' personal details at a large scale. 

Hence, we make use of a single-key FHE primitive such as BFV  
\cite{cryptoeprint:2012:144}, 
and we still need a (fully trusted) Central Server (CS) (for e.g., the ID issuing authority) to decrypt the encrypted data. But we significantly reduce the burden on CS by outsourcing most of the computations on encrypted data to TPS. 


\subsection{Our Contributions}

\begin{itemize}
    \item  We propose a practical privacy-preserving digital identity verification protocol involving SP, CS and TPS, using a single-key FHE scheme (see Figure \ref{fig:fig-1} and Section \ref{sec:ppid}).
    \item Our protocol supports matching of encrypted demographic and biometric data. Only the ID number of the user record is in the plaintext format to allow quick indexing by TPS to locate his/her encrypted data. We currently support exact match of all the demographic information, age comparison queries, and fuzzy fingerprint matching (see Section \ref{sec:qtypes}). 
    \item The main technical idea used in making our protocol efficient on the CS side is to design an ``extended and query independent'' FHE decryption circuit that decrypts intermediate encrypted values resulting from a low multiplicative-depth circuit evaluation on ciphertexts, thereby, outsourcing the rest of the computations to TPS and SP in a secure manner. 
    The extended part of the decryption circuit basically does the necessary Boolean operations on plaintexts very efficiently. One of the advantages of an extended decryption circuit being independent of the type of query is that it facilitates cheaper hardware implementation.  To facilitate this query transparent extended decryption, we propose a novel user data encoding mechanism.
Hence, our protocol is an important step towards fully outsourcing computations on encrypted user data  to TPS. Further, we believe that the encoding techniques we have proposed can
positively impact privacy-preserving protocols across many
applications.
	\item We give a complete security analysis of the proposed protocol in Section \ref{sec:security}.
    \item We have implemented our protocol using the Microsoft SEAL library \cite{sealcrypto} that implements the BFV FHE scheme. 
    The ciphertext of each individual user's data is only 0.864 MB (for 192-bit security level) as we deploy FHE ciphertext batching and it takes at most  0.04 seconds of computing time of TSP for demographic match queries, 0.3 seconds for age comparison queries, and 0.3 seconds for the biometric match query on a \textit{single} core (See Table \ref{Table:tab1}). The compute time for CS is less than 5 milliseconds per query. We would like to stress that the above timings correspond to \textit{latency}, not the amortised time as is typically reported in FHE implementations. The communication overhead is also very less due to the small size of data.  
    We would like to note that since our goal in this work is to outsource as much computations as possible to TPS, it would not be a fair comparison if we compare the above timings with prior works that all outsource only arithmetic operations to TPS. 
\end{itemize}

\ignore{
For future work, it will be interesting to design a digital ID verification protocol where the role of (a fully trusted) CS is only during the enrollment of users. Secure Multi-Party Computation (MPC)-based protocols could provide a practical solution to this problem assuming the existence of two or more non-colluding TPS (note that we need to minimize the role of CS, so it is best to avoid computation and communication load on it). Another practical concern in many developing countries is the availability of reliable Internet services. Hence, it will be useful to develop protocols that make very limited or no use of communication between entities, particularly, from/to SP. A solution for the latter problem could potentially make use of tamper-resistant hardware, thereby, increasing the cost of deployment and maintenance. 
} 


\section{Privacy-Preserving ID Verification Protocol}
\label{sec:ppid}
In this section, we describe the overall architecture of our protocol, including all the parties involved and their respective roles. We then describe the details of 
 user data encoding we propose, and the set of possible queries on users' data that our protocol can handle.
 We analyse the security of the proposed protocol in Section \ref{sec:security}. 
 The performance of our protocol is discussed in Section \ref{sec:expt}.

\subsection{System Architecture}
\begin{figure*}[h]


    \centering
    \includegraphics[width=\textwidth, scale = 0.6]{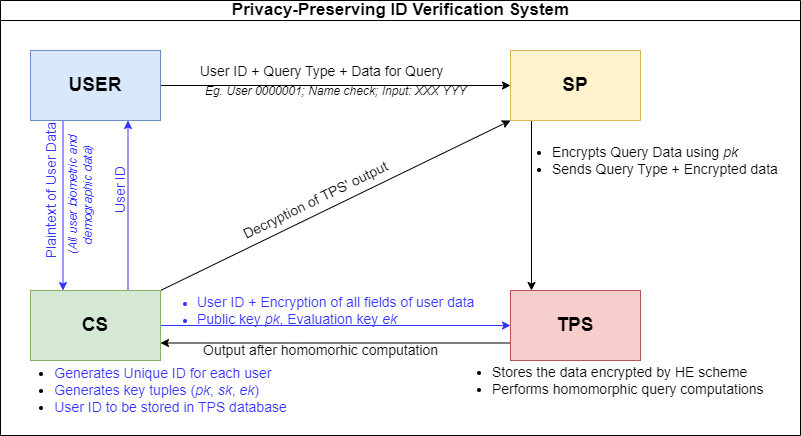}
    \caption{Flowchart of system architecture. The blue text/arrows represent the flow during initial user registration, and the black text/arrows represent the flow during query computation}    
    \vspace{2em}
    \label{fig:fig-1}
\end{figure*}

The following are the entities that are involved in the proposed system. Their roles and relationships are also summarised in Figure \ref{fig:fig-1}.

\begin{itemize}
    \item \textbf{Users}: 
    Users initially enroll in the digital ID system owned by CS by providing his/her demographic and biometric (fingerprint) information. At a later point of time he/she would avail the services of an SP that needs identity verification. 
    
    \item \textbf{Service Providers (SP)}: These are entities that require identity verification for providing a service. They can be government organisations, such as ration distribution centres or private companies such as telecom or gas providers. They collect users' demographic and biometric data and generate ID verification queries on this data, which in turn gets forwarded to TPS. 
    
    \item \textbf{Third-Party Servers (TPS)}: These are the third-party servers that may be located anywhere on earth. They provide storage and computation services. They receive queries from SPs, fetch the appropriate (encrypted) user records using the unique User ID numbers as the indices, and perform computations on the encrypted data itself. They then send the output of these encrypted matching computations to CS.
    
    \item \textbf{Central Servers (CS)}: These are the secure servers that will be owned by the governments of the nations deploying the ID system. As mentioned before, CS enrolls users into the ID system and securely stores the users' demographic and biometric data. Also, since the system deploys a single-public-key FHE scheme, CS servers are the only entities that will possess the FHE decryption key, and hence, will be the only point at which decryption may be carried out. So CS servers receive the encrypted query output from TPS and perform an ``extended'' decryption operation on it.
The extended decryption circuit performs the ciphertext decryption operation first and then passes the decrypted data through another circuit that determines whether a query has passed/failed, the output of which is forwarded to SP. By making our post decryption computations deterministic and query agnostic, these implementations can be made efficient, and require lesser hardware investment to setup.
\end{itemize}

\subsection{User Data and Encryption}
\label{sec:udata}
Users' data collected during enrollment is of two types: demographic and biometric. Demographic data includes user name, date-of-birth, gender, email, phone number, and pincode (see Table \ref{Table:tab1}), while the biometric data is the user's fingerprint information stored as fingercodes vector  \cite{fingercodes,fingercodes2}. The exact format in which each part of the user data is stored will be discussed in detail during the description of the corresponding query. Note that each user is issued a unique User ID number after enrollment that is stored as plaintext at TPS. 

In order to make computations on large ciphertexts more efficient, we use the \textit{CRT Batching} via SEAL's \texttt{BatchEncoder} class. Let $N$ denote the degree of the polynomial modulus and $T$ denote the plaintext modulus. Batching allows the BFV plaintext polynomials to be viewed as $2$-by-$(N/2)$ matrices, where each element is an integer modulo $T$. In the matrix view, encrypted operations act element-wise on encrypted matrices, allowing us to obtain speeds-ups of several orders of magnitude by making our computations fully vectorizable.
In the proposed system, $N$ is set to 8192, and $T$ is set to a 22-bit prime number. Our encrypted data is thus stored as two batched vectors of size 8192, one for demographic data and one for biometric (fingerprint) data. Table \ref{Table:tab1} summarises the data attributes stored for each user, its size in plaintext, the corresponding vector indices, query timings and the ciphertext size. Note that since many slots in the plaintext vectors are empty, we could add data for multiple fingerprints, or more detailed demographic data in the plaintext vectors for better utilisation  of storage in future implementations.

\begin{table*}[h]
\setlength{\tabcolsep}{5pt}
\centering

\begin{tabular}{|c|c|c|c|c|c|}
\hline
\textbf{Field}                                               & \textbf{\begin{tabular}[c]{@{}c@{}}Plaintext \\ Size \\ (bytes)\end{tabular}} & \textbf{\begin{tabular}[c]{@{}c@{}}Vector \\ Index \end{tabular}} & \textbf{\begin{tabular}[c]{@{}c@{}}Query Computation \\ Time at TPS (sec) \\ (latency)\end{tabular}} & \textbf{\begin{tabular}[c]{@{}c@{}}Storage \\ Format \end{tabular}}                                                            & \textbf{\begin{tabular}[c]{@{}c@{}}Size on \\ Disk at \\ TPS (KB)\end{tabular}} \\ \hline
User ID                                                      & 16                                                                         & -                     & -                                                                               & Plaintext                                                                          & 0.016                                                                 \\ \hline
Name                                                         & 50                                                                         & {[}0, 399{]}          & \multirow{5}{*}{0.04 (single thread)}                                       & \multirow{6}{*}{\begin{tabular}[c]{@{}c@{}} Encrypted\\ Demographic\\  Data Vector\\ $\text{demo}_\text{enc}$ \end{tabular}} & \multirow{6}{*}{432}                                                  \\ \cline{1-3}
Gender                                                       & 1                                                                          & {[}400, 407{]}        &                                                                                 &                                                                                    &                                                                       \\ \cline{1-3}
Pincode                                                      & 6                                                                          & {[}408, 455{]}        &                                                                                 &                                                                                    &                                                                       \\ \cline{1-3}
Phone Number                                                 & 13                                                                         & {[}456, 559{]}        &                                                                                 &                                                                                    &                                                                       \\ \cline{1-3}
Email ID                                                     & 20                                                                         & {[}560, 799{]}        &                                                                                 &                                                                                    &                                                                       \\ \cline{1-4}
Date-of-birth                                                & 6                                                                          & {[}800, 1599{]}       &  0.3 (single thread)                                                       &                                                                                    &                                                                       \\ \hline
\begin{tabular}[c]{@{}c@{}}Biometric\\ Template\end{tabular} & 640                                                                        & {[}0, 640{]}          &  0.3 (single thread)                                                         & \begin{tabular}[c]{@{}c@{}} Encrypted\\ Biometric\\ Data Vector\\ $\text{bio}_\text{enc}$  \end{tabular}                    & 432                                                                   \\ \hline
\end{tabular}
\vspace{0.5em}

\caption{Summary of the user data and the way it is stored in ciphertext. The timing data corresponds to a single threaded implementation on a 20-core Intel Xeon(R) Silver 4414 processor. The compute time for CS is less than 5 milliseconds per query.} 
\label{Table:tab1}
\end{table*}


\subsection{Query Types and Algorithms}
\label{sec:qtypes}
The following are the types of queries that can be currently addressed in our system. It is worth noting, however, that by using FHE, we can address new query types that may arise without any changes to our encrypted data.

\begin{enumerate}
    \item \textbf{Exact Matching of Demographic Data}: these are the queries that exactly match the demographic data, character by character. Since a query only passes after an exact match of personal details, these queries can be easily addressed with integer arithmetic.
    We store our demographic data in a single \textit{batch vector}, where each index of the vector stores 1 bit of data. Note that for each English character we can use an 8-bit encoding such as ASCII. The SEAL operations of homomorphic multiplication and vector rotation allow us to homomorphically isolate any part of this vector. Hence, for any demographic exact match query, TSP will isolate the appropriate part of the (encrypted) demographic data vector using a (query dependent) masked vector of 0s and 1s, and then the CS runs the next steps of our matching algorithm on the decrypted data. 
 
     For instance, to perform the exact match query on a user's pincode (which is stored in the index positions [408, 455] of the (encrypted) demographic vector $\text{demo}_\text{enc}$ ), we can homomorphically multiply the vector with the encryption of a vector of all 0s except in the index positions [408, 455], where it has a 1. Then, on left-shifting the resulting vector by 408 positions, we achieve the encryption of a vector having the pincode in index [0, 47]. 
     Let the vector created above be $v_{\text{enc}}$. Our matching algorithm works as follows:
     \begin{itemize}
         \item \textbf{At SP}: encode the input pincode to be verified in a vector, say $u$, in index positions [0, 47]. Encrypt this vector (call it $u_{\text{enc}}$) and send it to TPS along with the query type so that TPS knows which data is being verified.
        
         \item \textbf{At TPS}: Use the User ID (which is in the plaintext format) to locate the encrypted demographic record $\text{demo}_\text{enc}$  corresponding to this user that was previously outsourced from CS. Calculate $v_{\text{enc}}$ as indicated above. Then calculate $v_{\text{enc}} - u_{\text{enc}}$ homomorphically. Send the output to CS.
        
         \item \textbf{At CS}: Check if the output vector is $ \langle 0, 0, 0, \cdots, 0 \rangle$, i.e., the all-zero vector. If yes, then send ``Query Passed'' message to SP, else, send ``Query Failed''.
     \end{itemize}    

The above steps are formally described in Algorithm \ref{Algorithm:alg1}. 

    \begin{algorithm}[p!]
        \caption{Demographic data matching algorithm}
        \label{Algorithm:alg1}
        \begin{algorithmic}[1]
            \Algphase{At SP}
            \Require User data string $s$\Comment{Data here can be any of the demographic fields}
            \Procedure{EncodeData}{$s$}
                \State $u \gets \text{binary encoding of }s$
                \State $u_{\text{enc}} \gets \text{Enc}_{pk}(u)$
\Comment{$pk$ is the FHE public key of CS}                
                \State $u_{\text{enc}}$ is sent to the TPS as query data.
            \EndProcedure
            
            \Algphase{At TPS}
            \Require User ID $id$, encrypted input data vector $u_{\text{enc}}$, query type $q$ 
            \Function{CompareDemographic}{$u_{\text{enc}}, id, q$} 
\Comment{\parbox[t]{.4\linewidth}{\linespread{1}\selectfont Assume for the sake of description that $q$ is an email data}}            
            \State $\text{demo}_{\text{enc}} \gets \text{encrypted demographic data vector corresponding to } id \text{ from database}$
            \State $v_{\text{enc}} \gets \text{Enc}_{\textit{pk}}(\langle 0, 0, \cdots , 1, \cdots , 1, 0, \cdots , 0, \rangle$) \Comment{\parbox[t]{.4\linewidth}{\linespread{1}\selectfont Vector of all 0s, with 1s in the index positions $[560, 799]$, where the email address is stored in $\text{Dec}_{\textit{sk}}(\text{demo}_{\text{enc}})$}}
            \State $v_{\text{enc}} \gets v_{\text{enc}} \times \text{demo}_{\text{enc}}$ 
\Comment{\parbox[t]{.4\linewidth}{\linespread{1}\selectfont Isolating only the email data from $\text{demo}_{\text{enc}}$}}             
            \State $v_{\text{enc}} \gets $LeftShift($v_{\text{enc}}, 560$) \Comment{\parbox[t]{0.4\linewidth}{\linespread{1}\selectfont Shift encrypted email data from indices $[560, 799]$ to indices $[0, 239]$}}
            \State $\text{out}_{\text{enc}} \gets v_{\text{enc}} - u_{\text{enc}}$
            \State Send $\text{out}_{\text{enc}}$ to CS.
            \EndFunction
            \Algphase{At CS}
            \Require $\text{out}_{\text{enc}}$ from TPS, secret key \textit{sk}
            \Procedure{CheckOutput}{$\text{out}_{\text{enc}}$} \Comment{Procedure to check if query data was a match}
            

            \State $\text{out} \gets \text{Dec}_{\textit{sk}}(\text{out}_{\text{enc}})$
            \If{Indices $[0, 400]$ of $\text{out}$ are 0}
            \State \textbf{return} "Query data match successful!"
            \Else
            \State \textbf{return} "Query data match failed"
            \EndIf
            \EndProcedure
            
        \end{algorithmic}
    \end{algorithm}

Hence, we can process the exact demographic matching queries with a circuit having zero multiplicative depth as was known previously. Also, note that the proof of correctness of the above matching algorithm is straightforward.

    \item \textbf{Fuzzy Matching of Biometric Data}: Our protocol supports approximate matching of fingerprints. We record the fingerprint data as a fingercodes vector, which is a vector of size 640 bytes. This vector is encoded as a batch vector with each vector element representing 1 byte of data. 
    The fingercodes matching algorithm declares two fingerprints to be a match if the squared Euclidean distance between their corresponding vectors is less than or equal to a certain non-zero threshold $\beta$. 
    The improved fingercodes comparison algorithm we implement  achieves a genuine acceptance rate of $>99\%$ whilst limiting false acceptances to $<1\%$ as demonstrated in
\cite{fingercodes2}. 
    
    Having access to ciphertext addition, ciphertext multiplication, and ciphertext vector rotation operations, we can efficiently calculate the (squared) Euclidean distance between two BFV encrypted vectors (also true for any other FHE scheme that supports arithmetic operations and batching). This particular step is efficiently computed as suggested in \cite[Algorithm 1]{Boddeti18} and is detailed in Algorithm \ref{Algorithm:alg2}. In order to check if the value is less than the threshold $\beta$, we try to introduce a $0$ at some index position in the vector. 
The following computations are carried out:
     \begin{itemize}
        \item \textbf{At SP}: collect the fingerprint data from the user. Encode this as a fingercodes vector with the first 640 indices populated. Encrypt this vector to obtain the input test vector $u_{\text{enc}}$.         
         \item \textbf{At TPS}:
         \begin{itemize}
         	  \item Use the User ID to locate the encrypted biometric record $\text{bio}_{\text{enc}}$.
         	  \item Using ciphertext subtraction, squaring, and rotation and addition operations, homomorphically compute the squared Euclidean distance $\text{ED}$ between $u_{\text{enc}}$ and $\text{bio}_{\text{enc}}$. Let the resulting ciphertext $e_{\text{enc}}$ correspond to the data vector $\langle \text{ED}, 0, 0, \cdots, 0 \rangle$.        	  	
              \item Transform the vector $e_{\text{enc}}$ using rotation and addition operations to correspond to the data vector $ \langle \text{ED}, \text{ED}+1, \text{ED}+2, \cdots, \text{ED}+\beta, 0, 0, \cdots, 0 \rangle$.
             \item Create a ciphertext $b_{\text{enc}}$ corresponding to the data vector $ \langle \beta, \beta, \cdots, \beta \rangle$.
             \item Output the homomorphic subtraction $e_{\text{enc}} - b_{\text{enc}}$ to CS.
         \end{itemize}
         \item \textbf{At CS:} 
         Decrypt the ciphertext received from TPS. If it contains a 0 on any of its index positions, then send  ``Query Passed'' message to the SP, else send ``Query Failed''.
               
     \end{itemize}

\pagestyle{empty}

The above steps are formally described in Algorithm \ref{Algorithm:alg2}. 
    \begin{algorithm}[p!]
        \caption{Biometric data threshold comparison algorithm}
        \label{Algorithm:alg2}
        \begin{algorithmic}[1]
            \Algphase{At SP}
            \Procedure{EncodeFingerprintData}{}
                \State $u \gets \text{encoding of the user's fingerprint data}$ \Comment{\parbox[t]{.4\linewidth}{\linespread{1}\selectfont $u$ here is a fingercodes vector with the first 640 positions populated with 1 byte of data each.}}
                \State $u_{\text{enc}} \gets \text{Enc}_{pk}(u)$
                \State $u_{\text{enc}}$ is sent to the TPS as query data.
            \EndProcedure
            
            \Algphase{At TPS}
            \Require User ID $id$, encrypted biometric data vector $u_{\text{enc}}$, query type $q$, threshold value $\beta$ 
                
                    

            \Function{CompareBiometric}{$u_{\text{enc}}, id, q$} \Comment{$q$ here will be ``Biometric"}
            \State $v_{\text{enc}} \gets \text{encrypted biometric data vector corresponding to } id \text{ from database}$
                    \Procedure{CalculateEuclideanDistance}{$u_{\text{enc}}, \text{bio}_{\text{enc}}$}
                    \State $e_{\text{enc}} \gets u_{\text{enc}} - \text{bio}_{\text{enc}}$
                    \State $e_{\text{enc}} \gets e_{\text{enc}} \times e_{\text{enc}}$
                    \State $temp \gets e_{\text{enc}}$
                    \State $i \gets 0$
                    \While{$(i < 640)$}
                        \State $temp \gets \text{LeftShift}(temp, 1)$
                        \State $e_{\text{enc}} \gets e_{\text{enc}} + temp$
                        \State $i \gets i+1$
                    \EndWhile
                    \State $temp \gets \text{Enc}_{\textit{pk}}(\langle 1, 0, 0, \cdots , 0 \rangle)$ \Comment{\parbox[t]{0.4\linewidth}{\linespread{1}\selectfont Encryption of vector of all 0s except at index position 0}}
                    \State $e_{\text{enc}} \gets e_{\text{enc}} \times temp$ \Comment{\parbox[t]{0.4\linewidth}{\linespread{1}\selectfont $e_{\text{enc}}$ is now of the form $\langle \text{ED}, 0, 0, \cdots , 0 \rangle$, where ED is the Euclidean distance.}}
                \EndProcedure
                
                \State \hfill
                \Procedure{CompareThreshold}{$e_{\text{enc}}, \beta$}
                    \State $i \gets 0$
                    \State $temp \gets e_{\text{enc}}$
                        
                    
                        \While{($i \leq \beta$)}
                        \State $temp \gets \text{RightShift}(temp, 1)$
                        \State $e_{\text{enc}} \gets e_{\text{enc}} + temp$
                        \State $i \gets i+1$
                        \EndWhile
                   \State $e_{\text{enc}} \gets e_{\text{enc}} + \text{Enc}_{\textit{pk}}(\langle 0, 1, \cdots , \beta,0, \cdots\rangle)$                        \Comment{\parbox[t]{0.4\linewidth}{\linespread{1}\selectfont $e_{\text{enc}}$ is now of the form $\langle \text{ED}, \text{ED} + 1, \text{ED + 2}, \cdots , \text{ED} + \beta, 0, 0, \cdots , 0 \rangle$}}
                            \State $b_{\text{enc}} \gets \text{Enc}_{\textit{pk}}(\langle \beta, \beta, \cdots , \beta \rangle)$
                            \State $\text{out}_{\text{enc}} \gets e_{\text{enc}} - b_{\text{enc}}$
                        \EndProcedure
         \State Send $\text{out}_{\text{enc}}$ to CS.
                        
                        
                    
                    \EndFunction

                    \Algphase{At CS}
                    \Require $\text{out}_{\text{enc}}$ from TPS, secret-key \textit{sk}
                    \Procedure{CheckOutput}{$\text{out}_{\text{enc}}$} \Comment{\parbox[t]{0.4\linewidth}{\linespread{1}\selectfont Procedure to check is query data was a match}}
                    \State $\text{out} \gets \text{Dec}_{\textit{sk}}(\text{out}_{\text{enc}})$
                    \If{ $\text{out}$ contains one 0}
                    \State \textbf{return} "Query data match successful!"
                    \Else
                    \State \textbf{return} "Query data match failed"
                    \EndIf
                    \EndProcedure
                        \end{algorithmic}
                    \end{algorithm}

Below we reason why the above procedure leads to correctly computing the comparison $\text{ED} \leq \beta$. The intuition is that if $\text{ED} \leq\beta$, then $\exists x$, such that $x \leq \beta$ and $\text{ED}+x-\beta = 0$. 
          
\textbf{Proof of correctness:} As per the steps at TPS, the encrypted vector $e_{\text{enc}}$ will correspond to the plaintext vector $\langle \text{ED}, \text{ED} + 1, \text{ED + 2}, \cdots , \text{ED} + \beta, 0, 0, \cdots , 0 \rangle$, where $\text{ED}$ is the squared Euclidean distance between the test and the original biometric vectors. As per the fingercodes algorithm, the two fingerprints may be declared a match if $\text{ED} \leq \beta$, where $\beta$ is the specified threshold value. We know that if $\text{ED} \leq \beta$, then $(\text{ED} + x) - \beta = 0$, where $x \in [0, \cdots, \beta]$. Hence, we introduce $0$ in this vector by calculating $e_{\text{enc}} - b_{\text{enc}}$, where recall that $b_{\text{enc}}$ is the encryption of $\langle \beta, \beta, \cdots , \beta \rangle$. The resulting vector will be $\langle \text{ED} - \beta, \text{ED}+1-\beta, \text{ED}+2-\beta, \cdots , \text{ED}+\beta-\beta, T-\beta, T-\beta, \cdots , T-\beta \rangle$, where $T$ is the chosen plaintext modulus in the BFV FHE scheme. Since $\beta$ is non-zero, $T-\beta$ will also be non-zero. Therefore, at CS, if the decrypted vector contains a 0, then we know that $\text{ED} \leq \beta$. 
        
    
    \item \textbf{Age Comparison}: Many services require age verification. We need our system to be able to answer age comparison queries on FHE encrypted ciphertexts. For e.g., ``Is the user above the age of 18?'' or ``Is the user below the age of 65?''. This requires us to create a comparison algorithm using arithmetic circuits, which are natively supported by the BFV scheme. We formulate these query types in a more generic manner as follows:
    
    \begin{quote}
        Given two dates $d_1$ and $d_2$, can we encode them as vectors and perform a set of operations on these vectors such that the resulting vector has properties that only hold when $d_1$ comes before $d_2$?
    \end{quote}
    We encode our date-of-birth data as distance from a pivot date that we set it as 1 Jan, 1900 (this date can be changed if needed). At a high level, our algorithm involves using logic gates to determine which of $d_1$ and $d_2$ is closer to our pivot year (1900). In case both $d_1$ and $d_2$ fall in the same year, we use the distance from the pivot day (1 Jan) to determine which of them comes first.
We would like to stress that this encoding of dates is done in the same data vector where other demographic information is stored.\\ 
 
\pagestyle{plain}
   
\textbf{Simulating Logic Gates: AND, NOT, OR, XOR}.   
     Our solution to the age comparison problem requires the use of logic gates on batch data vectors. Given two encrypted batch vectors whose underlying data vectors $v_1$ and $v_2$ are such that each element of these vectors is either 0 or 1, and suppose we want to evaluate logic gates such as AND, NOT, OR and XOR on these binary sequences. It is easy to see that an AND gate is the same as (homomorphically) calculating the product of two vectors ($0 \times 1 = 0$, and $1 \times 1 = 1$). However, addition will not be a successful substitute for an OR gate, as $1 + 1 = 2$, which is not a binary value. 
    
     Our approach to this was to come up with an algorithtm to simulate the NOT gate, which would then give us access to a NAND gate, a universal logic gate. Given a \textit{binary} vector $v$, our algorithm for computing $\bar{v}$ (i.e., NOT $v$) is as follows:
     \begin{itemize}
     \renewcommand{\labelitemi}{$-$}
         \item Encode a vector, say $u$, with each of its elements being $T - 1$ ($T$ is the plaintext modulus), i.e., $u = \langle T-1, T-1, \cdots, T-1 \rangle$. Since BVF operates in modular arithmetic, this is equivalent to having each element set to -1.
         \item Encode a vector, say $w$, with each of its elements being 1, i.e., $w = \langle 1, 1, 1, \cdots, 1 \rangle$. 
         \item (Homomorphically) evaluate $w + (u \times v)$. This is the required vector $\bar{v}$.
     \end{itemize}
We can simulate OR and XOR gates using the following well-known logic circuits:
    \begin{equation*}
        \begin{split}
            v_1 \vee v_2 &= \overline{\overline{v_1} \wedge \overline{v_2}}, \\
            v_1 \oplus v_2 &= (v_1 \vee v_2) \wedge \overline{v_1 \wedge v_2}. \\
        \end{split}
    \end{equation*}
Hence, we can homomorphically evaluate the OR and the XOR gates on BFV batched ciphertexts using circuits having multiplicative depths 3 and 4, respectively.

     Now we explain how the date-of-birth (DoB), or  in general, any date is encoded into the demographic data vector:
 \begin{itemize}
     \item We pick a pivot date of 1 Jan, 1900. The given date is then expressed as distance from this pivot in years and days. For instance, the date 6 April, 1999, can be viewed as being 99 years after 1900, and 96 days after 1 Jan. So it can be expressed as the pair (099, 096). In this way, every date can be expressed as the number of years from 1900, $y$, and the number of days from 1 Jan, $d$.
    
     \item In our demographic data vector, we use 800 index positions to store this date at index positions [800, 1599]. Of these, the first 400 positions (i.e., indices [800, 1199]) contain $y$ encoded in unary, and the remaining 400 positions (i.e., indices [1200, 1599]) store the vector $\langle d, d+1, d+2, \cdots, d+399 \rangle$. For instance, 6 April, 1999, is encoded as:
     \begin{equation*}
         \langle \underbrace{1, 1, 1, \cdots, 1}_\text{first 98 indices}, \underbrace{0, 0, \cdots 0,}_\text{next 302 indices} 96, 97, 98, \cdots, 495 \rangle.
     \end{equation*}
     \end{itemize}
     During query execution at TPS, we can isolate each of the two parts of this encrypted vector using a technique  similar to direct demographic comparison. We create vectors $y_{\text{enc}}$ and $d_{\text{enc}}$ such that:
    \begin{equation*}
        \begin{split}
            y_{\text{enc}} &= \text{Enc}(\langle1, 1, 1, \cdots, 1, 0, 0, \cdots, 0 \rangle) \\
            d_{\text{enc}} &= \text{Enc}(\langle96, 97, 98, \cdots, 494, 495 \rangle)
        \end{split}
    \end{equation*}
    from the demographic vector $\text{demo}_{\text{enc}}$ corresponding to the User ID.
    
     To compare this date to some input date, we create the corresponding vectors at SP $y'_{\text{enc}}$ and $d'_{\text{enc}}$ for this input date. Here, $y'_{\text{enc}}$ is created similar to $y_{\text{enc}}$, but $d'_{\text{enc}}$ does not increment as in $d_{\text{enc}}$, i.e., if the number of days from 1 Jan for the input date-of-birth are $x$, then $d'_{\text{enc}}$ is constructed as $\text{Enc}(\langle x, x, x, \cdots, x \rangle)$.
    
     Now, given $y_{\text{enc}}$, $d_{\text{enc}}$, $y'_{\text{enc}}$ and $d'_{\text{enc}}$, each being an encrypted vector with data in index positions [0, 399], we execute the following algorithm at TPS homomorphically:
     
     \textbf{At TPS}:
     \begin{itemize}
     	 \item Use the User ID to locate the encrypted demographic record $\text{demo}_{\text{enc}}$. Compute ciphertexts $y_{\text{enc}}$ and $d_{\text{enc}}$ from $\text{demo}_{\text{enc}}$ as explained above. Receive $y'_{\text{enc}}$ and $d'_{\text{enc}}$ from SP.
         \item Define $\text{temp}_1 = y_{\text{enc}} \wedge (y_{\text{enc}} \oplus y'_{\text{enc}}) $
         \item Define $\text{temp}_2 = (d'_{\text{enc}}-d_{\text{enc}}) \times \overline{(y_{\text{enc}} \oplus y'_{\text{enc}}) \wedge y'_{\text{enc}}} $
         \item  $\text{temp}_2 \gets \text{RightShift}(\text{temp}_2, 400)$ 
         \item $\text{out}_{\text{enc}} \gets \text{temp}_1 + \text{temp}_2$.  This is the final output.
     \end{itemize}
     After the above sequence of operations at TPS, $\text{out}_{\text{enc}}$ is decrypted at CS to, say, vector $w$. CS next does the following:
     \begin{quote}
         The input (test) date comes (strictly) after the user's DoB \textit{iff} at indices [0, 399] of $w$ contain only 0's, and indices [400, 799] of $w$ contain at least one 0. In this case, ``Query Passed'' message is sent to SP.  Else, the input date becomes on or before the user's DoB. In the latter case, ``Query Failed'' is sent.
     \end{quote}

Note that to answer queries of the form \textit{is age greater than 18?}, to the actual DoB 18 years must be added to obtain the date corresponding to the user turning 18. This can be easily computed homomorphically at TPS.      

The above steps are formally described in Algorithm \ref{Algorithm:alg3}.

    \begin{algorithm}[p!]
        \caption{Date-of-birth comparison algorithm}
        \label{Algorithm:alg3}
        \begin{algorithmic}[1]
            \Algphase{At SP}
            \Require User date-of-birth $dob$
            \Procedure{EncodeDateofBirth}{$dob$}
                \State $y \gets \text{no. of years from 1900 in }dob$
                \State $d \gets \text{no. of days from 1st January in }dob$
                \State ${y'}_{\text{enc}} \gets \text{Enc}_{pk}(\langle \underbrace{1, 1, \cdots, 1,}_{\text{index}\,[0, y-1]} 0, 0, \cdots , 0\rangle)$ \Comment{\parbox[t]{0.34\linewidth}{\linespread{1}\selectfont $y$ no. of 1s, followed by all 0s upto index 400}}
                \State ${d'}_{\text{enc}} \gets \text{Enc}_{pk}(\langle \underbrace{d, d, \cdots, d,}_{\text{index}\,[0, 399]} 0, 0, \cdots , 0\rangle)$ \Comment{Vector of all $d$'s upto index 399}
                \State ${y'}_{\text{enc}}$ and ${d'}_{\text{enc}}$ are sent to TPS as query data.
            \EndProcedure
            
            \Algphase{At TPS}
            \Require User ID $id$, ${y'}_{\text{enc}}$ and ${d'}_{\text{enc}}$ from SP. 
            \Function{CompareDateOfBirth}{$id, {y'}_{\text{enc}}, {d'}_{\text{enc}}$} 
                \State $\text{demo}_{\text{enc}} \gets \text{encrypted demographic data vector corresponding to } id \text{ from database}$
                \State \Comment{\parbox[t]{0.4\linewidth}{\linespread{1}\selectfont The year and date in $\text{demo}_{\text{enc}}$ are stored in indices [800,1199] and [1200,1599] respectively.}} 
                
                \State $y_{\text{enc}} \gets \langle 0, 0, \cdots, 0, \underbrace{1, 1, \cdots, 1,}_\text{index [800, 1199]} 0, \cdots, 0 \rangle$
                    \State $d_{\text{enc}} \gets \langle 0, 0, \cdots, 0, \underbrace{1, 1, \cdots, 1,}_\text{index [1200, 1599]} 0, \cdots, 0 \rangle$
                        \State $y_{\text{enc}} \gets \text{LeftShift}(\text{demo}_{\text{enc}} \times y_{\text{enc}}, 800)$
                \State $d_{\text{enc}} \gets \text{LeftShift}(\text{demo}_{\text{enc}} \times d_{\text{enc}}, 1200)$
                \State \hfill
                \State $\text{temp}_1 \gets y_{\text{enc}} \wedge (y_{\text{enc}} \oplus {y'}_{\text{enc}})$ \Comment{\parbox[t]{0.4\linewidth}{\linespread{1}\selectfont All logic gates used are homomorphic and element-wise across vectors}}
                
                \State $\text{temp}_2 \gets ({d'}_{\text{enc}} - d_{\text{enc}}) \times \overline{(y_{\text{enc}} \oplus {y'}_{\text{enc}}) \wedge {y'}_{\text{enc}}}$
                \State $\text{temp}_2 \gets \text{RightShift}(\text{temp}_2, 400)$
                \State $\text{out}_{\text{enc}} \gets \text{temp}_1 + \text{temp}_2$
                
                
                
                \State $\textbf{return } \text{out}_{\text{enc}} \text{ to CS}$
                
            \EndFunction
            
            \Algphase{At CS}
            \Require $\text{out}_{\text{enc}}$ from TPS, secret key \textit{sk}
            \Procedure{CheckOutput}{$\text{out}_{\text{enc}}$} \Comment{Procedure to check if input DoB lies after user's DoB in the database}
            \State $\text{out} \gets \text{Dec}_{\textit{sk}}(\text{out}_{\text{enc}})$
            \If{ Index $[0, 399] \text{ of }\text{out}$ are all $0$s and index $[400, 799]$ of out contain at least one $0$}
            \State \textbf{return} "Input DoB lies after the user's DoB"
            \Else
            \State \textbf{return} "Input DoB lies before the user's DoB"
            \EndIf
            \EndProcedure
            \end{algorithmic}
        \end{algorithm}

Below we argue about the correctness of the proposed age/date comparison algorithm.

        \textbf{Proof of Correctness:} Let $y$ and $d$ be the values of the distance from the year and date pivots for the user's DoB, and $y'$ and $d'$ be these distances for the input date. Let $y_{\text{enc}}, d_{\text{enc}}, y'_{\text{enc}}, d'_{\text{enc}}$ be the corresponding encrypted vector encodings as described previously. We can categorise various scenarios into four cases.
    
    \begin{itemize}
        \item \textbf{Case 1:} $y > y'$
        
    
    
 Note that the data vector corresponding to $\text{temp}_1$ contains 1s if $y$ has more 1s than $y'$. In other words, the first 400 index positions of the output vector contain a non-zero index only if the input date comes before the user's DoB. 
Since the first 400 index positions will not be 0s, 
the output will declare input date to come before the user's DoB. Hence, due to the check on $\text{temp}_1$, the algorithm works correctly in all cases when $y > y'$.

    \item \textbf{Case 2:} $y = y'$ and $d > d'$
    
    When $y$ and $y'$ are equal, $y_{\text{enc}}$ and $y'_{\text{enc}}$ will correspond to the same underlying data vector. Hence $y_{\text{enc}} \oplus y'_{\text{enc}} = \text{Enc}(\langle 0, 0, \cdots, 0 \rangle)$. 
    Since $y = y'$, then $\overline{(y_{\textit{enc}} \oplus {y'}_{\text{enc}}) \wedge {y'}_{\text{enc}}}$ will be an encryption of $\langle 1, 1, \cdots, 1 \rangle$. The data corresponding to the vector $\text{temp}_2$ will therefore be the same as $d'_{\text{enc}} - d_{\text{enc}}$.
    We know that if $d > d'$, then $d'-(d+i)$ will be non-zero for all positive integers $i$. Hence, none of the indices of the plaintext vector corresponding to $d'_{\text{enc}} - d_{\textit{enc}}$ will be zero. 
    The output will, therefore, be correct in all cases when $y = y'$ and $d > d'$.
    
    \item \textbf{Case 3:} $y = y'$ and $d \leq d'$
    
    As discussed in Case 2, $y_{\text{enc}} \oplus y'_{\text{enc}}$ will be an encryption of zero, and the output will be determined by $\text{temp}_2 = d'_{\text{enc}} - d_{\text{enc}}$. 
    We know that if $d < d'$, then $\exists$ $i \geq 0$ such that $d' - (d + i) = 0$. Also, $i$ will be smaller than the maximum possible value of $d'$. The vector $d_{\text{enc}}$ encodes the value $d+i$ in each of its indices, with $0 \leq i < 400$. Since the maximum possible distance any date can have from 1 Jan is 366 (number of days in a leap year) and $366 < 400$, we know that exactly one index of the data vector corresponding to $d'_{\text{enc}} - d_{\text{enc}}$ will contain a 0 when $d < d'$. 
    The first 400 indices of the output vector will be all 0s ($y = y'$), and the next 400 indices will contain at least one 0. The algorithm will declare the user's DoB to lie before the input date. Hence, the algorithm output is correct in all cases when $y = y'$ and $d \leq d'$.

    \item \textbf{Case 4:} $y < y'$
    
    
     Since $y \neq y'$, as in Case 1, $y_{\text{enc}} \oplus y'_{\text{enc}}$ will contain 1s in the index positions $[y, y'-1]$. However, since the extra 1s are in $y'_{\text{enc}}$, $\text{temp}_1$ 
     will be:
     \begin{align*}
         \text{temp}_1 & = y_{\text{enc}} \wedge (y_{\text{enc}} \oplus y'_{\text{enc}})\\ 
         & = \text{Enc}(\langle 0, 0, 0, \cdots, 0, 0 \rangle).
     \end{align*}
     In Cases 2 and 3, the value corresponding to $\text{temp}_2$ was independent of that corresponding to $\overline{(y_{\text{enc}} \oplus y'_{\text{enc}}) \wedge  y'_{\text{enc}} }$, since this has been a vector of all 1s. 
     However, in this case, since the number of 1s in $y_{\text{enc}}$ is less than the number of 1s in $y'_{\text{enc}}$, the vector $\overline{(y_{\text{enc}} \oplus y'_{\text{enc}}) \wedge y'_{\text{enc}} }$ will be of the form:
     \begin{align*}
         \text{Enc}(\langle \underbrace{1, 1, \cdots, 1}_{\text{Index} [0,y-1]}, 0, 0, 0, 0, 0, \underbrace{1, 1, \cdots, 1}_{\text{Index} [y', 399]}  \rangle). 
     \end{align*}
     The value of the second half of the output vector will be
     $\text{temp}_2 = $
     \begin{equation*}
     \begin{split}
          \text{Enc}(\langle \underbrace{X, X, \cdots, X}_{\text{Index} [0,y-1]}, 0, 0, 0, 0, 0, \underbrace{X, X, \cdots, X}_{\text{Index} [y', 399]}  \rangle).
     \end{split}
     \end{equation*}
     Here, $X$ is the value of that index in the vector $d'_{\text{enc}} - d_{\text{enc}}$, which may not be 0. By multiplying this vector by $\overline{  (y_{\text{enc}} \oplus y'_{\text{enc}}) \wedge y'_{\text{enc}}}$, we introduce 0s in the second half of our output vector whenever $y < y'$. Therefore, irrespective of the values of $d$ and $d'$, the output vector will contain at least one 0 in the index positions [400, 799] every time $y < y'$. 
 Hence, the output will be correct in all cases when $y < y'$.

    \end{itemize}
    
    The above-described cases document the age comparison procedure's 
     behaviour for all possible values of $y, d, y'$ and $d'$. Hence, its correctness is proved.

     \item \textbf{Central Server - Query Agnostic Processing}: 

Next, we explain how the computations done at CS after  decryption of the ciphertext are query agnostic. 
    Let us recap below the query PASS condition, as determined by CS, for each algorithm, given $v$, the decrypted vector output from TPS:
    \begin{itemize}
        \item \textbf{Demographic Comparison}: Query PASS if indices [0, 399] of $v$ are all 0s.
        \item \textbf{Biometric Comparison}: Query PASS if indices [0, $\beta-1$] of $v$ contain at least one 0 ($\beta$ here is the fingercodes threshold).
        \item \textbf{Date-of-birth Comparison}: Query PASS if indices [0, 399] of $v$ are all 0s, and indices [400, 799] contain at least one 0. (\textit{Note}: PASS case here is the input date lying after the user's DoB)
    \end{itemize}
    
    In the configuration of MS SEAL used in this project, the size of each vector is 4096, and the size of the fingercodes threshold $\beta$ is 3000. Therefore, we can carry out homomorphic operations at TPS on the final vector $v_{\text{enc}}$ to create a modified vector $v'_{\text{enc}}$ for each algorithm as follows: 
    \begin{itemize}
        \item \textbf{Demographic Comparison}: 
        \[
        v'_{\text{enc}} = v_{\text{enc}} \times \text{Enc}(\langle \underbrace{1, 1, \cdots, 1}_{\text{Index [0, 399]}}, \underbrace{0, 0, \cdots, 0}_{\text{Index [400, 400+$\beta$]}} \rangle)
        \]
        \item \textbf{Biometric Comparison}: $v'_{\text{enc}} = \text{RightShift}(v_{\text{enc}}, 400)$
        \item \textbf{Date Comparison}: $v'_{\text{enc}} = v_{\text{enc}} + \text{Enc}(\langle \underbrace{0, 0, \cdots, 0}_{\text{Index [0, 799]}}, \underbrace{1, 1, \cdots, 1}_{\text{Index [800, 400 + $\beta$]}} \rangle)$
    \end{itemize}
    
    By performing these computations at TPS before sending the final vector (now $v'$) to CS, the following check can be used for all query types at CS:
    \begin{quote}
        If indices [0, 400] of $v'$ are all 0s and indices [400, 400+$\beta$] contain at least one 0, then send to SP ``Query Passed''. Else send ``Query Failed''.
    \end{quote}
     In this way, our post decryption check at CS is the same for all query types. 
    
\end{enumerate}


\section{Security Guarantees}
\label{sec:security}

In this section, we provide a security analysis of the proposed protocol in Fig. \ref{fig:fig-1}. 

\subsection{Security Model}

Recall that the entities other than the users in the protocol are the CS, TSP, and SPs. The privacy objective of our protocol is to protect the privacy of the users' demographic and biometric data.

The CS is assumed to be fully trusted as only this entity possesses the decryption key. It is still preferable for CS to maintain a backup of user data in encrypted form to safeguard against malware attacks (though there could still be leakage during decryption). 
By making the decryption of intermediate computations available only to CS, which is trusted, our protocol is resistant to large-scale user data leakage attacks. For future work, it is interesting to explore the side-channel/fault attack protection of the decryption operation at CS.

Our protocol assumes that the SPs too are honest, and it does \textit{not} collude with TSP. It is important to ensure that SP does not collude with TPS because SP would have access to the user data as a plaintext (that needs to be verified) and in a typical scenario it would correspond to a genuine user. This is also the reason that the SP is assumed to be trusted. Note that SP would additionally learn only the result of ID verification. Extra care must be taken when deploying this protocol in practice to make sure that the nature of SP's business is not leaked to TPS when the former makes an ID verification query. For instance, if the TPS gets to know that SP is the pension disbursal authority, then it will be able to conclude that the age of the queried user is beyond the retirement age. Such a scenario can be avoided by providing SPs with an identifier. 

We provably claim below in Section \ref{sec:proof} that our protocol is secure against an honest-but-curious adversary controlling TPS. 

\textbf{Leakage at TPS through User ID}:
because the User ID is stored in the plaintext format for efficient indexing by TPS, it will be able to track where all and when ID verification request for this user was done, and also the type of the ID verification query. But, due to encryption, we show that it will not be able to determine any other information about the user.

Note that the final verification result must be directly sent from CS to SP as proposed in our protocol. Else, using this information the TPS may be able to successfully brute force over all possible guesses to eventually determine the correct user data.   

\subsection{Security Proof}
\label{sec:proof}

\begin{lemma}
An honest-but-curious adversary controlling TPS will not learn anything about users' personal data in the protocol in Fig. \ref{fig:fig-1}.
\end{lemma}

Let $\mathcal{T}$ denote the transcript of the protocol corresponding to TPS. That is, $\mathcal{T}$ will contain all the messages sent to and received from  TPS in the complete execution of the protocol. Let  
\[
\mathcal{T} = \mathcal{T}_1 \cup \mathcal{T}_2,
\]   
where  
\[
\mathcal{T}_1 = \{\; (\text{UserID}^{(i)}, \text{demo}^{(i)}_{\text{enc}}, \text{bio}^{(i)}_{\text{enc}})\; \}\qquad \forall i. 
\]
Hence, $\mathcal{T}_1$ contains the UserIDs and the corresponding (homomorphic) ciphertexts of the demographic and biometric encodings of every user in the system.
The set $\mathcal{T}_2$ contains all the messages received by TPS during the query execution and the encrypted query processed outputs sent to CS. The input consiste of the user IDs and the corresponding ciphertexts of the demographic and biometric encodings to be verified that was sent by SP. That is,
\[
\mathcal{T}_2 = \{\; (\text{UserID}^{(j)}, \text{QueryType}^{(j)}, \text{demo}'^{(j)}_{\text{enc}}, \text{bio}'^{(j)}_{\text{enc}},\text{output}^{(j)}_{\text{enc}})\; \}\qquad \forall j. 
\]  

We next prove the existence of an efficient \textit{Simulator} $\mathcal{S}$ that can simulate the transcript $\mathcal{T}^I$  in the \textit{ideal} world that is indistinguishable from the \textit{real} world transcript  $\mathcal{T}$ defined above. In the ideal world, the Simulator $\mathcal{S}$ is only given the following inputs
\[
\mathcal{W}_1 = \{\; (\text{UserID}^{(i)})\; \}\qquad \forall i, 
\]
and
\[
\mathcal{W}_2 = \{\; (\text{UserID}^{(j)},\text{QueryType}^{(j)})\; \}\qquad \forall j. 
\]  
Then $\mathcal{S}$ simulates the following transcript
\[
\mathcal{T}^I_1 = \{\; (\text{UserID}^{(i)}, \text{RandDemo}^{(i)}_{\text{enc}}, \text{RandBio}^{(i)}_{\text{enc}})\; \}\qquad \forall i, 
\]
where $\text{RandDemo}^{(i)}_{\text{enc}}$ and $\text{RandBio}^{(i)}_{\text{enc}}$ are ciphertexts corresponding to the demographic and biometric encodings of the user data chosen \textit{uniform randomly} from the permitted data ranges. Since the underlying FHE scheme is CPA secure, it readily follows that
\[
\mathcal{T}^I_1 \approx \mathcal{T}_1,
\]
that is, the above two distributions are computationally indistinguishable. Similarly, the transcript 
\[
\mathcal{T}^I_2 = \{\; (\text{UserID}^{(j)}, \text{QueryType}^{(j)}, \text{RandDemo}'^{(j)}_{\text{enc}}, \text{RandBio}'^{(j)}_{\text{enc}},\text{output}'^{(j)}_{\text{enc}})\; \}\qquad \forall j 
\]  
is constructed on the lines of $\mathcal{T}^I_1$. Note that $\text{RandDemo}'^{(j)}_{\text{enc}}$ and $\text{RandBio}'^{(j)}_{\text{enc}}$ are again encryptions of \textit{randomly} chosen user data, and $\text{output}'^{(j)}_{\text{enc}}$ corresponds to the homomorphic query processing as would be done in the real world with ``original'' data for a given user taken instead from $\mathcal{T}^I_1$. Again, from the CPA security of the underlying FHE scheme, it follows that 
\[
\mathcal{T}^I_2 \approx \mathcal{T}_2.
\]
Let $ \mathcal{T}^I = \mathcal{T}^I_1 \cup \mathcal{T}^I_2,$ then we have   
\[
\mathcal{T}^I \approx \mathcal{T}.
\]
Hence, the Simulator $\mathcal{S}$ is able to simulate an ideal world transcript of TPS that is computationally indistinguishable from the real world transcript of TPS. This completes the security analysis. 

\section{Experimental Results}
\label{sec:expt}

As described in Section \ref{sec:udata}, we choose our polynomial modulus degree as 8192, and our plaintext modulus $T$ as a 22 bit prime in the SEAL FHE library. It automatically picks the remaining parameters to achieve 192-bit security by default, which can be changed to 128-bit or 256-bit as needed. The third-party server uses $<$ 1MB of storage for each user's data (from a dummy data set).
Each message sent from SP to TPS, and from TPS to CS is a ciphertext of size 432 KB (see \Cref{Table:tab1}). 

In our biometric comparison algorithm, we implement optimisations as suggested in \cite[Algorithm 1]{Boddeti18}  to speed up the algorithm. In the loops at lines 13 and 25 of Algorithm \ref{Algorithm:alg2}, we use the fast exponentiation trick to complete the loop operations in $\lceil \log_2(640) \rceil$ and $\lceil \log_2(\beta) \rceil$ steps, respectively. 

We run our experiments on a laptop with an Intel i7-7700HQ processor running at 2.8 GHz. All operations are run as \textit{single-threaded} tasks. For our experiments, we execute each query 1000 times and use the C++ \texttt{chrono} class to collect timing data (latency) in microseconds. The average time taken over 1000 iterations for each query is summarised in Table \ref{tab:timings}.


\begin{table}[htb!]
\caption{Timing data of different query types at TPS and decryption at CS. In the fingercodes implementation used, the parameter $\beta$ is set to 3000.}
\begin{tabular}{|c|c|c|c|c|}
\hline
\textbf{Operation}                                                      & \textbf{\begin{tabular}[c]{@{}c@{}}Multiplicative\\ Depth\end{tabular}} & \textbf{\begin{tabular}[c]{@{}c@{}}No. of\\ Rotations\end{tabular}}                                   & \textbf{\begin{tabular}[c]{@{}c@{}}No. of\\ Additions\end{tabular}}                                       & \textbf{\begin{tabular}[c]{@{}c@{}}Time\\ (milliseconds) \\ (latency) \end{tabular}} \\ \hline
CS: decryption                                                          & -                                                                       & -                                                                                                     & -                                                                                                         & 4.66                                                                   \\ \hline
Name match                                                              & \multirow{5}{*}{1}                                                      & 1                                                                                                     & \multirow{5}{*}{1}                                                                                        & 22.82                                                                  \\ \cline{1-1} \cline{3-3} \cline{5-5} 
Gender match                                                            &                                                                         & \multirow{4}{*}{2}                                                                                    &                                                                                                           & 35.39                                                                  \\ \cline{1-1} \cline{5-5} 
Pincode match                                                           &                                                                         &                                                                                                       &                                                                                                           & 40.83                                                                  \\ \cline{1-1} \cline{5-5} 
Phone Number match                                                      &                                                                         &                                                                                                       &                                                                                                           & 35.36                                                                  \\ \cline{1-1} \cline{5-5} 
Email ID match                                                          &                                                                         &                                                                                                       &                                                                                                           & 35.17                                                                  \\ \hline
\begin{tabular}[c]{@{}c@{}}Date-of-birth \\ comparison\end{tabular}     & 7                                                                       & 3                                                                                                     & 6                                                                                                         & 217.73                                                                 \\ \hline
\begin{tabular}[c]{@{}c@{}}Biometric\\ Template comparison\end{tabular} & 3                                                                       & \begin{tabular}[c]{@{}c@{}}$\lceil \log_2(640) \rceil$\\ $+ \lceil \log_2(\beta) \rceil$\end{tabular} & \begin{tabular}[c]{@{}c@{}}$\lceil \log_2(640) \rceil$\\ $+ \lceil \log_2(\beta) \rceil + 2$\end{tabular} & 286.74                                                                 \\ \hline
\end{tabular}
\label{tab:timings}
\end{table}

As the results indicate, our most compute-intensive query takes $< 0.3$ to execute, making it perfectly feasible for real-world deployment. The timing information does \textit{not} account for network latency. We can expect a further speedup in these timings by using Intel's Homomorphic Encryption Acceleration Library (HEXL) \cite{hexl} along with an AVX512-IFMA52 compatible processor.


%
%

\section{Future Work}





For future work, it will be interesting to design a digital ID verification protocol where the role of (a fully trusted) CS is only during the enrollment of users. Secure Multi-Party Computation (MPC)-based protocols could provide a practical solution to this problem assuming the existence of two or more non-colluding TPS (note that we need to minimise the role of CS, so it best to avoid computation and communication load on it). Another practical concern in many developing countries is the availability of reliable Internet services. Hence, it will be useful to develop protocols that make very limited or no use of communication between entities, particularly, from/to SP. A solution for the latter problem could potentially make use of tamper-resistant hardware, thereby, increasing the cost of deployment and maintenance. We could also explore extending our protocol to accommodate a malicious TPS. Finally, it will be interesting to explore other applications for the user data encoding scheme we have proposed.

\ifnum\ANON=0

\section*{Acknowledgements}
This work was funded in part by the MOSIP project and by the Infosys Foundation Career Development Chair Professorship grant for the second author. We would like to thank Shyam Murthy (IIIT Bangalore), Ramesh Narayanan (MOSIP), Sanath Kumar (MOSIP), and Sasikumar Ganesan (MOSIP) for valuable technical discussions.

\ignore{

\section*{Speaker's biography} Dr. Srinivas Vivek is currently the Infosys Foundation Career Development chair professor at IIIT Bangalore. He was a postdoctoral researcher in the Cryptography group at the University of Bristol, UK. He has obtained his doctoral degree from the University of Luxembourg (Luxembourg). He has served as a member of the editorial board/PC of IACR Trans. CHES, IACR Communications in Cryptology,  CT-RSA, CARDIS, WAHC, AsianHOST, Indocrypt, ICISS, etc. He is also a recipient of the DST INSPIRE faculty award from Govt. of India. 
His research interests are in the design, analysis and implementation of (1) countermeasures against side-channel attacks, and (2) privacy-preserving protocols using homomorphic encryption schemes.  
}

\fi

\bibliographystyle{plain}
\bibliography{ref}

\end{document}